\documentstyle[epsfig,12pt]{article}

\input amssym.def       
\input amssym.tex

\def\double{\Bbb}

\def\ccc{{\double C}}     
\def\nnn{{\double N}}       
\def\zzz{{\double Z}}
\def\rrr{{\double R}}

\def\aa{{\cal A}}
\def\bb{{\cal B}}

\def\dd{{\cal D}}

\def\hh{{\cal H}}
   
\def\mm{{{\cal M}}}

\def\t{\mathrm{Tr}} 

\def\dix{\int \!\!\!\!\!\! -}

\def\lp{\left(} 
\def\rp{\right)}
\def\la{\left\{} 
\def\ra{\right\}}

\def\ul{\underline}
\def\ov{\overline}
\def\ot{\otimes}
\def\op{\oplus}

\def\bbb{\begin{eqnarray}}
\def\eee{\end{eqnarray}}
\def\bbbb{\begin{eqnarray}}
\def\eeee{\end{eqnarray}}

\def\pp{\pmatrix}
\def\n{\nonumber}

\newtheorem{thm}{Theorem}[section]

\newtheorem{dfi}{Definition}[section]

\newtheorem{pro}{Proposition}[section]

\def\demo{\noindent\ul{Proof}:\\ \noindent }
\def\edemo{\hfill$\square$\\}

\begin{document}

\vskip 2truecm
\begin{center}
{\bf\Large \sc Gauge invariance of the Chern-Simons action\\
\medskip
in noncommutative geometry}
\end{center}
\bigskip
\begin{center}
{\bf Thomas KRAJEWSKI}
\footnote{ and Universit\'e de Provence and Ecole Normale Superieure de Lyon, tkrajews@cpt.univ-mrs.fr} \\
CENTRE DE PHYSIQUE THEORIQUE\\
CNRS - Luminy, Case 907\\
13288 Marseille Cedex 9
\end{center}
\bigskip

\centerline{\bf Abstract} 
\medskip
In complete analogy with the classical case, we define the Chern-Simons action functional in noncommutative geometry and study its properties under gauge transformations. As usual, the latter are related to the connectedness of the group of gauge transformations. We establish this result by making use of the coupling between cyclic cohomology and K-theory and prove, using an index theorem, that this coupling is quantized in the case of the noncommutative torus. 



\section{Introduction}

It is nowadays well admited that the major difficulty encountered in developping a quantum theory of gravity lies in our current conception of geometry. Indeed, it is known that if we want to observe a particle localized in a very small region of space-time of size $\Delta x$, we have to use an other particle with momentum $\Delta p$ such that $\Delta x \Delta p \geq \hbar$. Thus, the smaller the uncertainty on the position is, the larger the momentum $\Delta p$ is, so that compatibilty with Einstein's equations shows that a singularity appears in the limit $\Delta p\rightarrow 0$ \cite{dop}.

\par

As a consequence, one has to give up the standard notion of localization in space-time itself and not only in phase space, as taught by quantum mechanics. To proceed, one may impose non trivial commutation relations between the space-time coordinates in order to obtain suitable uncertainty relations. The development of geometrical concepts within this context leads us directly to {\it noncommutative geometry}, which may be defined as the geometry of spaces whose coordinates fail to commute.
 
\par

This new area of mathematics ranges from operators algebras to quantum groups, the latter appearing, for instance, when one tries to describe the symmetries of the quantum plane, whose coordinates satisfy $xy=qyx$ with $q\in\ccc^{*}$. Here, we will be mainly interested in the theory developped by A. Connes which relies on the use of operator algebraic concepts \cite{bible}. Roughly speaking, it may be summed up by the two following steps. First, one tries to formulate a geometrical theory like measure theory or topology using a suitable subalgebra of the algebra of complex valued functions on the space $X$ under consideration. Then one extends the previous theory to more general algebras that are not necesarily commutative; these algebras are to be thought as algebras of coordinates on the "quantum space" replacing $X$. These algebras $\aa$ are always subalgebras of the algebras of  bounded operators on a given Hilbert space $\hh$; for instance Von Neumann algebras and $C^{*}$-algebras are respectively relevant in the study of the noncommutative extension of measure theory and topology.  

\par

However, the notions relevant to physics are of differential nature, so that one must develop differential geometry in the noncommutative setting.
Borrowing ideas from quantum mechanics, one replaces the derivative of a function by a commutator with a suitable operator $\dd$ acting on $\hh$, so that one may define the differential of a "function" $a\in\aa$ by $da =[\dd,\pi(a)]$, $\pi$ being a representation of $\aa$ as operators on $\hh$. Accordingly, it turns out that the relevant notion is that of a {\it spectral triple} $(\aa,\hh,\dd)$ which is supposed to satisfy additional requirements given below. In the commutative case, if we assume that $\mm$ is a compact manifold endowed with a spin structure, one can reconstruct all differential geometric notions from the spectral triple $(\aa,\hh,\dd)$, where $\aa$ is the algebra of smooth function on $\mm$ represented by multiplication on the Hilbert space of square integrable sections of the spinor bundle and $\dd$ is the standard Dirac operator.

\par

In the general case, starting with such a triple one can reconstruct the analogue of gauge theory, even with nontrivial topological properties. Furthermore, one can built in all these cases a Yang-Mills action functional which exhibits all standard properties of a bona fide Yang-Mills action: positivity, gauge invariance, etc... Here, our main concern will be the construction of the Chern-Simons action, which has proved to be relevant, in the classical case, to many areas of mathematics and physics. Because of the nontrivial properties of this action under gauge transformation, we will have to use the machinery of noncommutative geometry including the coupling of cyclic cohomology to K-theory \cite{ihes} and the index theorem \cite{local}.

\section{Spectral triples and differential forms}

To construct differential geometric objects in noncommutative geometry, the relevant notion is that of a spectral triple that we already have introduced above. Let us now precise its definition \cite{bible}.

\begin{dfi}
A spectral triple $(\aa,\hh,\dd)$ consists in an involutive algebra $\aa$ together with a faithfull representation $\pi$ of $\aa$ by bounded operators on a separable Hilbert space $\hh$. $\dd$ is an unbounded self-adjoint operator with compact resolvent and such that $[\dd,\pi(a)]$ is bounded for any $a\in\aa$.
\end{dfi}

Furthermore, one may assume that the triple $(\aa,\hh,\dd)$ satisfies additional requirements stated in order to recover spin geometry from commutative spectral triples \cite{grav}. Amongst all these  requirements commonly refered to as "axioms of noncommutative geoemtry", we will only make use of the following two.

\bigskip

\noindent
{\bf Axiom 1 (Dimension)}
\begin{it}
There is a positive integer $n$ such that the decreasing sequence $(\lambda_{k})_{k\in\nnn}$ of the eigenvalues of the compact operator $ds=|\dd|^{-1}$ satisfies
\bbb
\lambda_{k}=O\lp \frac{1}{k^{1/n}}\rp
\eee
when $k\rightarrow \infty$.
\end{it}

\bigskip

This axiom only gives us a lower bound on $n$, but together with the other axioms it defines the dimension of a spectral triple. Here, by dimension of a spectral triple we esentially mean an integer satisfying the previous axiom.

\par

Since $\dd$ has compact resolvent, its kernel is finite dimensional and $ds=|\dd|^{-1}$, with $|\dd|=\sqrt{D^{2}}$ is well defined on the orthogonal complement of $\ker\dd$. This axiom implies that the sequence of eigenvalues of $ds^{n}=|\dd|^{-n}$ is bounded by the sequence $C/k$ for $C>0$ and $k$ large enough, so that the Dixmier trace $\t_{\omega}(T|\dd|^{-n})$ is well defined for any bounded operator $T$ \cite{bible}. In the commutative case one can recover the ordinary integral of a function from the Dixmier trace since we have 
\bbb
\t_{\omega}\lp \pi(f)|\dd|^{-n}\rp=
\frac{2^{n-[n/2]}}{\pi^{n/2}\Gamma(n/2+1)}\int_{\mm}\,d^{n}x\sqrt{g}\,f,
\eee
for any smooth function $f$ on a compact Riemannian manifold of dimension $n$. Accordingly, we define for any spectral triple of dimension $n$ and any bounded operator $T$ the analogue of the integral
\bbb
\dix{T}ds^{n}=\frac{\pi^{n/2}\Gamma(n/2+1)}{2^{[n/2]-n}}\t_{\omega}\lp T|\dd|^{-n}\rp,
\eee
bearing in mind that all operators in the algebra generated by $\pi(\aa)$ and $[\dd,\pi(\aa)]$ are bounded operators. 

\par

The other axiom we will need is the regularity axiom.

\bigskip

\noindent
{\bf Axiom 3 (Regularity)}
\begin{it}
Any element $b$ of the algebra generated by $\pi(\aa)$ and $[\dd,\pi(\aa)]$ lies in the domains of the powers of the derivation defined by $\delta(b)=[|\dd|,b]$.  
\end{it}

\bigskip

{}From this axiom, one deduces the following result \cite{cipriani}.

\begin{pro}
Let $\bb$ be the algebra generated by $\pi(\aa)$ and $[\dd,\pi(\aa)]$. Then the map $b\mapsto \t_{\omega}\lp b|\dd|^{-n}\rp$ is a trace on $\bb$.
\end{pro}  
\bigskip

This trace property proves to be of primary importance when we study the behavior of the action functional under gauge invariance. 

\par

Let us now tackle the question of the construction of  differential forms. We first introduce a formal construction called {\it universal differential algebra}. 

\begin{dfi}
Let $\aa$ be a unital algebra. The universal differential algebra over $\aa$ is the graded algebra  $\Omega(\aa)=\op_{k\in\nnn}\,\Omega^{k}(\aa)$, where $\Omega^{k}(\aa)$ is the vector space generated by
\bbb
a_{0}\delta a_{1}\dots\delta a_{k}
\eee
for any $a_{0},a_{1},\dots,a_{k}\in\aa$. The product is obtained by simple juxtaposition together with the relations  $\delta(ab)=\delta a\,b+a\,\delta b$ for any $a,b\in\aa$ as well as $\delta(1)=0$. The exterior derivative $d\,:\, \Omega^{k}(\aa)\rightarrow \Omega^{k+1}(\aa)$ is the linear map defined by
\bbb
d\lp a_{0}\delta a_{1}\dots\delta a_{k}\rp=
\delta a_{0}\delta a_{1}\dots\delta a_{k}
\eee
for all $a_{0},a_{1},\dots,a_{k}\in\aa$.
\end{dfi}

It fulfils the standard properties of a differential algebra.

\begin{pro}
The exterior derivative $d$ is nilpotent and fulfils the graded Leibniz rule  $d(\omega\xi)=d\omega\,\xi+(-1)^{p}\omega\,d\xi$ for all $\omega\in\Omega^{p}(\aa)$ and $\xi\in\Omega^{q}(\aa)$.
\end{pro}

However, this construction is a rather formal one and has to be represented at the level of the Hilbert space by replacing the derivative by a commutator.  Accordingly, we define a representation of $\Omega(\aa)$ by
\bbb
\pi\lp a_{0}\delta a_{1}\dots\delta a_{k}\rp=
\pi(a_{0})[\dd,\pi(a_{1})]\dots[\dd,\pi(a_{k})].
\eee
Although this map defines a representation of $\Omega(\aa)$ as an algebra, it fails to be a representation of the differential structure. Indeed, this requires that we define the differrential of $\pi(\omega)$ as $d\pi(\omega)=\pi(d\omega)$ for any $\omega\in\Omega(\aa)$, which is not possible as soon as $\ker\pi\neq\la 0\ra$. 

\par

To proceed, let us define the ideal $J=\ker\pi+d\lp\ker\pi\rp$. The image of the quotient $\Omega(\aa)/J$ admits a well defined differential structure \cite{bible}. 

\begin{pro}
The graded algebra $\Omega_{\dd}(\aa)$ defined by
\bbb
\Omega_{\dd}(\aa)=\pi\lp \Omega(\aa)/J\rp=\pi\lp\Omega(\aa)\rp/\pi\lp d\ker\pi\rp  
\eee
admits an exterior derivative $d$ such that $d\pi(\omega)$ is a representative of the class defined by $\pi(d\omega)$, for any $\omega\in \Omega(\aa)$. This exterior derivative is nilpotent and satisfies the graded Leibniz rule. 
\end{pro} 

Unfortunately, we are now dealing with equivalence classes that usually admit more than one representative. In the simple case of a spectral triple obtained by tensoring the ordinary geometry of space-time by a matrix algebra, one can define a scalar product on $\pi(\Omega(\aa))$ by
\bbb
\langle \pi(\omega),\pi(\eta)\rangle=\t_{\omega}\lp \pi(\omega)^{*}\pi(\eta)|\dd|^{-n}\rp
\eee 
for any $\omega,\eta\in\Omega^{k}(\aa)$, whereas forms of different degree are defined to be orthogonal \cite{zylinski}. Let us notice that $\pi(\omega)$ and $\pi(\eta)$ are bounded operators, so that the previous expression is well defined. However, it is not in general a scalar product because we cannot check that it is positive definite. Indeed, it could happen that the eigenvalues of $\pi(\omega)$ decrease sufficiently fastly so that the trace vanishes, even if $\pi(\omega)\neq 0$. 

\par

Since we are interested in defining the Chern-Simons action, we need a three dimensional spectral triple. Accordingly, we set $n=3$ from now on but the discussion of what follows may generalized to other values of $n$.

\par

The following condition defines the noncommuative analogue of a manifold without a boundary.

\begin{dfi}
A three dimensional spectral triple $(\aa,\hh,\dd)$ is said to satisfy the closedness condition if
\bbb
\t_{\omega}\lp [\dd,\pi(a_{0})]\dots[\dd,\pi(a_{3})]|\dd|^{-3}\rp=0
\eee
for any $a_{0},\dots,a_{3}\in\aa$.
\end{dfi}

As a computational device, it allows us to use the rule of integration by parts,
\bbb
\dix\lp\pi(d\omega)\pi(\eta)\rp ds^{3}=
(-1)^{p+1}\dix\lp\pi(\omega)\pi(d\eta)\rp ds^{3}
\eee
for any $\omega\in\Omega^{p}(\aa)$ and $\eta\in\Omega^{q}(\aa)$ with $p+q=3$.

\par

For later puposes, it is useful to relate this condition to cyclic and Hochschild cohomology. For completeness, we recall the following basic definitions \cite{loday}.  

\begin{pro}
Let $\aa$ an algebra and $\phi;\;\aa^{n+1}\rightarrow\ccc$ a (n+1)-linear map.
$\phi$ is said to be a Hochschild cocyle if it satisfies
\bbb
\mathop{\sum}\limits_{i=0}^{n-1}(-1)^{i}
\phi(a_{0},\dots,a_{i}a_{i+1},\dots,a_{n})
+(-1)^{n}\phi(a_{n}a_{0},a_{1},\dots,a_{n-1})=0
\eee
for any $a_{0},\dots,a_{n}\in\aa$. If in addition it fulfils
\bbb
\phi(a_{0},a_{1},\dots,a_{n})=(-1)^{n}\phi(a_{1},\dots,a_{n},a_{0}),
\eee
it is a cyclic cocycle.
\end{pro}

There is an easy caracterization of spectral triples fulfiling the closedness condition using cyclic cocycles.

\begin{pro}
A spectral triple $(\aa,\hh,\dd)$ satisfies the closedness condition if and only if the map $\phi:\; \aa^{4}\rightarrow\ccc$ defined by
\bbb
\phi(a_{0},a_{1},a_{2},a_{3})=
\t_{\omega}\lp \pi(a_{0})[\dd,\pi(a_{1})][\dd,\pi(a_{2})][\dd,\pi(a_{3})]|\dd|^{-3}\rp
\eee
is a cyclic cocycle.
\end{pro}

Obviously, if $(\aa,\hh,\dd)$ is a spectral triple, then $(M_{N}(\aa),\hh\ot\ccc^{N},\dd\ot I_{N})$ is a spectral triple satisfying all requirements imposed to  $(\aa,\hh,\dd)$. Moreover the former fulfils the closedness condition if only if the latter does. {}From now on, we shall work with the second one, whose differential forms are matrix valued forms.  

\par

Before applying this rather formal machinery to the construction of the Chern-Simons functional, let us adopt integral notations,
\bbb
\dix \t\lp \pi(\omega)\rp ds^{n}=\frac{\pi^{n/2}\Gamma(n/2+1)}{2^{n-[n/2]}}\t_{\omega}\lp\pi(\omega)|\dd|^{-n}\rp
\eee
for any matrix valued form $\omega\in M_{N}(\Omega(\aa))$,
as well as $da=[\dd,\pi(a)]$ for any $a\in M_{N}(\aa)$.

\section{Gauge invariance of the Chern-Simons action}

Before we come to grips with the noncommutative case, let us briefly recall some basic facts about Chern-Simons field theory \cite{witten}. If $\mm$ is a compact and orientable three dimensional manifold and $G$ is a compact Lie group which may be chosen to be $SU(N)$ for concreteness, the Chern-Simons action is defined to be
\bbb
S_{CS}[A]=\frac{k}{4\pi}\int\t\lp AdA+\frac{2}{3}A^{3}\rp,
\eee
where $k\in\rrr$ is a coupling constant and $A$ is a 1-form with values in the Lie algebra of $G$. It is a remarkable fact that this action does not depend on a metric on $\mm$ because we integrate a 3-form in dimension 3. This turns Chern-simons theory into a topological field theory \cite{blau}, whose quantization yields non trivial topological invariants of the manifold $\mm$ and allows us to recover the Jones polynomial of knot theory.

\par

Under the gauge transformation given by the map $g$ from $\mm$ into $G$, the gauge potential $A$ becomes $A^{g}=gAg^{-1}+gdg^{-1}$ and it is easy to show, using the standard properties of differential forms, that the Chern-Simons action is not gauge invariant,
\bbb
S_{CS}[A^{g}]=S_{CS}[A]+\frac{k}{12\pi}\int\t\lp gdg^{-1}\rp^{3}.
\eee
If we normalize the generators $T^{a}$ of the Lie algebra of $G$ such that  $\t(T^{a}T^{b})=-1/2\,\delta^{ab}$, one has
\bbb
\frac{1}{24\pi^{2}}\int\t\lp gdg^{-1}\rp^{3}=n
\eee
where $n$ is an integer called "winding number" of the map $g$ from $\mm$ into $G$.

\par

Accordingly, if $k$ is an integer, $e^{ikS_{CS}[A]}$ is gauge invariant and the partition function
\bbb
Z[\mm]=\int [\dd A]\,e^{ikS_{CS}[A]}
\eee
is well defined as a gauge theory. It is worthwhile to notice that the quantization of the coupling constant is preserved in the one-loop analysis because it is just shifted by an another integer \cite{witten}. 

\par

Furthermore, this integer has a deep topological significance because it is a measure of the defect of connectedness of the group of gauge transformations. Indeed, if $g_{0}$ and $g_{1}$ are two gauge transformations connected by a path $t\in[0,1]\mapsto g_{t}$, one can show that
\bbb
\frac{d}{dt}\int\t\lp g_{t}dg^{-1}_{t}\rp^{3}=0.
\eee
Accordingly, the winding number is constant on each connected component of the group of gauge transformations.

\par

Let us now come to the noncommutative case.

\begin{dfi}
Let $\lp\aa,\hh,\dd\rp$ be a spectral triple of dimension 3 satisfying the closedness condition and let $A\in M_{N}(\Omega_{\dd}^{1}(\aa))$ a hermitian  matrix of 1-forms. We define the Chern-Simons action as
\bbb
S_{CS}[A]=\dix\t\lp K_{1}\rp ds^{3},
\eee
where $K_{1}\in M_{N}(\pi(\Omega^{3}(\aa)))$ is any representative of the class of the Chern-Simons form $K=AdA+\frac{2}{3}A^{3}\in M_{N}\lp\Omega^{3}_{\dd}(\aa)\rp$.
\end{dfi}

To ensure self-consistency of this definition, we have to show that it only depends on the equivalence class of the Chern-Simons. If $K_{2}$ denotes an other representative of $K$, then, using the ideal properties of $J$, one has $\t(K_{1})-\t(K_{2})\in \pi(J)\cap\pi(\Omega^{3}(\aa))$ so that  
\bbb
\t\lp K_{1}\rp -\t\lp K_{2}\rp =\mathop{\sum}\limits_{i}
da_{1}^{i}da_{2}^{i}da_{3}^{i},
\eee  
where $a_{1}^{i}$, $a_{2}^{i}$ and $a_{3}^{i}$ are elements of $\aa$. {}From the closedness condition, we deduce that
\bbb
\dix\lp da_{1}^{i}da_{2}^{i}da_{3}^{i}\rp ds^{3}=0, 
\eee
so that
\bbb
\dix\t\lp K_{1}\rp ds^{3}=\dix\t\lp K_{2}\rp ds^{3}. 
\eee

\par

At first sight, it is not clear whether this action is of topological nature or not. Indeed, if it was of topological nature, it should only depend on the choice of the Dirac operator in a weak form, because the latter also contains information pertaining to the metric structure. Anyway, it is easy to see that in the commutative case one recovers the standard Chern-Simons action, which is doubtless of topological nature.  

\par

Let us now the study the gauge invariance of this action.

\begin{thm}
Let $\lp\aa,\hh,\dd\rp$ be a spectral triple of dimension 3 satisfying the closedness condition. Then, under the gauge transformation determined by a unitary element $u$ of $M_{N}(\aa)$, the Chern-Simons action becomes
\bbb
S_{CS}[uAu+udu^{-1}]=S_{CS}[A]+\Gamma[u],
\eee
with
\bbb
\Gamma[u]=-\frac{1}{3}\int\t\lp udu^{-1}udu^{-1}udu^{-1}\rp ds^{3}.
\eee
\end{thm}

\demo

To proceed, let us introduce the curvature $F=dA+A^{2}$ of $A$. We have
\bbb
AdA+\frac{2}{3}A^{3}=AF-\frac{1}{3}A^{3},
\eee
so that, if $F_{1}$ is a representative of $F$, the Chern-Simons action reads
\bbb
S_{CS}[A]=\dix\t\lp AF_{1}-\frac{1}{3}A^{3}\rp ds^{3}.
\eee
Under a gauge transformation, $A$ becomes $uAu^{-1}+udu^{-1}$ and $F$ transforms into $uFu^{-1}$. Using the ideal structure of $J$, it is clear that $uF_{1}u^{-1}$ is a representative of $uFu^{-1}$ and we have
\bbb
S_{CS}[uAu^{-1}+udu^{-1}]=\dix\t\lp \lp uAu^{-1}+udu^{-1}\rp uF_{1}u^{-1}
-1/3\lp uAu^{-1}+udu^{-1}\rp^{3}\rp ds^{3}.
\eee
Using the trace properties of the map  $\pi(\omega)\mapsto\dix\t(\pi(\omega))ds^{3}$ and  the relation $du^{-1}u+u^{-1}du=0$, we get 
\bbb
\dix\t\lp \lp uAu^{-1}+udu^{-1}\rp uF_{1}u^{-1}\rp ds^{3}=\dix\t\lp AF\rp ds^{3}-
\dix\t \lp duFu^{-1}\rp ds^{3},
\eee 
as well as
\bbbb
&-1/3\displaystyle\dix\t\lp uAu^{-1}+udu^{-1}\rp^{3} ds^{3}=-\frac{1}{3}\displaystyle\dix\t\lp A\rp^{3} ds^{3}&\n\\
&-\frac{1}{3}\displaystyle\dix\t\lp udu^{-1}\rp^{3} ds^{3}-\displaystyle\dix\t\lp udu^{-1}uA^{2}u^{-1}\rp ds^{3}+\displaystyle\dix\t\lp duAdu^{-1}\rp ds^{3}.&
\eeee
Gathering all terms, we obtain
\bbbb
&S_{CS}[uAu+udu^{-1}]=S_{CS}[A]+\Gamma[u]&\n\\
&+\dix\t\lp udu^{-1}uF_{1}u^{-1}-udu^{-1}uA^{2}u^{-1}+du Adu^{-1}\rp ds^{3}.&
\eeee
The operator appearing on the left hand side is a representative of the 3-form
\bbbb
udu^{-1}uFu^{-1}-udu^{-1}uA^{2}u^{-1}+du Adu^{-1}&=&udu^{-1}udAu^{-1}+duAdu^{-1}\n\\
&=&-dudAu^{-1}+duAdu^{-1}\n\\
&=&-dud\lp Au^{-1}\rp\n\\
&=&-d\lp ud(Au^{-1})\rp.
\eeee
Because this form is exact, the integral of all its representatives vanishes by the closedness condition,
\bbb
\dix\t\lp udu^{-1}uF_{1}u^{-1}-udu^{-1}uA^{2}u^{-1}+du Adu^{-1}\rp ds^{3}=0,
\eee
which proves that
\bbb
S_{CS}[uAu+udu^{-1}]=S_{CS}[A]+\Gamma[u].
\eee
\edemo

Let us now try to understand the topological significance of $\Gamma[u]$. We first have to recall the definition of the group $K_{1}(\aa)$.

\begin{dfi}
Let $\aa$ be a $C^{*}$-algebra (i.e. it is an involutive Banach algebra whose norm satisfies $||aa^{*}||=||a||^{2}$ for any $a\in\aa$) and let us denote by $U_{N}(\aa)$ the group of unitary elements of $M_{N}(\aa)$. Then, using the embedding
\bbb
u\in U_{N}(\aa)\mapsto \pp{u&0\cr 0&1}\in U_{N+1}(\aa)
\eee
we define
\bbb
U_{\infty}(\aa)=\mathop{\cup}_{N=1}^{\infty}U_{N}(\aa).
\eee
By definition, $K_{1}(\aa)$ is the group $\pi_{0}\lp U_{\infty}(\aa)\rp$ of connected components of $U_{\infty}(\aa)$.
\end{dfi}

We refer to \cite{wegge} for a general introduction to K-theory. It is important to point out that this definition works at the level of $C^{*}$-algebras, which corresponds to continuous functions, where the algebra $\aa$ appearing in the spectral triple $(\aa,\hh,\dd)$ forms the analogue of smooth functions. However, one can show, using holomorphic functional calculus, that this does not really matter \cite{ihes}. Thus, we work with $\aa$ as if it was a $C^{*}$-algebra.  

\begin{pro}
With the assumptions of the previous theorem, $\Gamma[u]$  only depends on the class of $u$ in the group $K_{1}(\aa)=\pi_{0}(U_{\infty}(\aa))$.
\end{pro}

\demo

This result relies on the coupling of cyclic cohomology to $K^{1}(\aa)$ \cite{ihes}: if $\phi_{2m+1}$ is an odd dimensional cyclic cocycle and $u$ unitary, then $\phi_{2m+1}(u-1,u^{-1}-1,\dots,u-1,u^{-1}-1)$ only depends on the class of $u$ in $K_{1}(\aa)$.

\par

Thanks to the closedness condition, $\Phi$ defined by
\bbb
\Phi(a_{0},a_{1},a_{2},a_{3})=\dix\lp a_{0}da_{1}da_{2}da_{3}\rp ds^{3},
\eee
is a cyclic cocycle on $\aa$ (cf proposition 2.5) that we extend to a cyclic cocyle $\tilde{\Phi}$ on $M_{N}(\aa)$ using the trace by
\bbb
\tilde{\Phi}(a_{0},a_{1},a_{2},a_{3})=\dix\t\lp a_{0}da_{1}da_{2}da_{3}\rp ds^{3}.
\eee
Consequently, $\tilde{\Phi}(u-1,u^{-1}-1,u,u^{-1}-1)$ only depends on the class of $u$ in $K_{0}(\aa)$. 

\par

Finally, let us notice that $\Gamma[u]=1/3\tilde{\Phi}(u,u^{-1}-1,u-1,u^{-1}-1)$ because of
\bbb
\lp udu^{-1}\rp^{3}=-udu^{-1}dudu^{-1}=-ud\lp u^{-1}-1\rp d\lp u-1\rp d\lp u^{-1}-1\rp.
\eee
Using the closedness condition, we get
\bbb
\tilde{\Phi}(1,u^{-1},u,u^{-1})=\dix\t\lp 1,u^{-1},u-1,u^{-1}-1\rp ds^{3}=0,
\eee
which proves that
\bbb
\Gamma[u]=\frac{1}{3}\tilde{\Phi}(u-1,u^{-1}-1,u-1,u^{-1}-1).
\eee
Accordingly, $\Gamma[u]$ only depends on the class of $u$. 
\edemo

Therefore, it is clear that $\Gamma[u]$ is constant on the connected components of $U_{\infty}(\aa)$ and the non invariance of the Chern-Simons functional is due to the defect of connectedness of the group $U_{N}(\aa)$ of gauge transformations, as in the classical case. 

\par

Although we prove that $\Gamma[u]$ only depends on the connected component in which $u$ lies, we have not shown that it is an integer up to a multiplicative constant. To proceed further, we have to relate it to the index of a given Fredholm operator.

\section{An application of the index theorem}

In the general case, it is always possible to associate to a spectral triple $(\aa,\hh,\dd)$ a class of Fredholm operators (i.e. it is a bounded operator with finite dimensional kernel and cokernel) whose index can be computed using a local formula involving cyclic cocycles \cite{local}. To state this result, we need the following definition.

\begin{dfi}
Let $(\aa,\hh,\dd)$ a spectral triple of dimension $n$ and let us denote by $\delta$ the derivation $\delta(b)=[|\dd|,b]$ for any $b$ in the algebra generated by $\pi(\aa)$ and $[\dd,\pi(\aa)]$. Then $(\aa,\hh,\dd)$ is said to have discrete dimension spectrum if there is a discrete subset $\Sigma\in\ccc$ such that the functions 
\bbb
\zeta_{b}(z)=\t(b|\dd|^{z}),
\eee
holomorphic for $\Re(z)$ large enough, extend holomorphicaly to $\ccc - \Sigma$ for any $b$ belonging to the algebra generated by the elements of $\bb$ and their images through $\delta^{k}$.
\end{dfi}

When $(\aa,\hh,\dd)$ has discrete dimension spectrum, we define on the algebra generated by  $\bb$ and $|\dd|^{z}$, $z\in\ccc$, a sequence $(\tau_{k})_{k\in\nnn}$ of linear functional by
\bbb
\tau_{k}(b)=\mathop{\mathrm{res}}\limits_{z=0} z^{k}\t\lp b|\dd|^{-2z}\rp.
\eee
In general, these functionals fail to be traces. However, when all the poles are simple, only $\tau_{0}$ is nontrivial and one can show that it is a trace which generalizes to the noncommutative case the Wodzicki residue.

\par

Let us now consider a spectral triple of dimension 3 and let us define on $\aa$ two cochains by 
\bbb
\phi_{3}(a_{0},a_{1},a_{2},a_{3})=
\frac{1}{12}\tau_{0}\lp a_{0}da_{1}da_{2}da_{3}]|\dd|^{-3}\rp
-\frac{1}{6}\tau_{1}\lp a_{0}da_{1}da_{2}da_{3}|\dd|^{-3}\rp
\eee
and
\bbbb
\phi_{1}(a_{1},a_{2})&=&
\tau_{0}\lp a_{0}da_{1}\rp\|dd|^{-1})
-\frac{1}{4}\tau_{0}\lp a_{1}\nabla(da_{1})|\dd|^{-3}\rp\n\\
&-&
\frac{1}{2}\tau_{1}\lp a_{1}\nabla(da_{1})|\dd|^{-3}\rp
+\frac{1}{8}\tau_{0}\lp a_{1}\nabla^{2}(da_{1})|\dd|^{-5}\rp\n\\
&+& \frac{1}{3}\tau_{1}\lp a_{1}\nabla^{2}(da_{1})|\dd|^{-5}\rp
+\frac{1}{12}\tau_{2}\lp a_{1}\nabla^{2}(da_{1}]|\dd|^{-5}\rp .\n
\eeee
where we have used $\nabla(b)=[\dd^{2},b]$ for any $b\in\bb$ and we write $da=[\dd,\pi(a)]$ and $a$ instead of $\pi(a)$ for any $a\in\aa$.

\par

Let us also introduce the unitary operator $F$ defined by $\dd=|\dd|F$ on the orthogonal complement of the finite dimensional kernel of $\dd$, as well as  its positive part $P=\frac{!+F}{2}$. Of course, we extend this construction to the spectral triple $(M_{N}(\aa),\hh\ot\ccc^{N},\dd\ot I_{N})$ so that we can deal with matrices over $\aa$.

\par

In dimension 3, we can formulate the index theorem as follows \cite{local}.
 
\begin{thm}
If $u\in M_{N}(\aa)$ is unitary, then $PuP$ is a Fredholm operator on $P\hh$  whose index is given by 
\bbb
\mathrm{Ind}(PuP)=
\phi_{1}(u,u^{-1})-\phi_{3}(u,u^{-1},u,u^{-1}).
\eee
\end{thm}

Since $\mathrm{Ind}(PuP)=\dim\ker PuP-\dim\ker Pu^{*}P$ is an integer, we always have
\bbb
\phi_{1}(u,u^{-1})-\phi_{3}(u,u^{-1},u,u^{-1})\in\zzz.
\eee
Unfortunately, even in the case of simple spectrum, only $\phi_{3}$ can be related to $\Gamma[u]$ and thus the index theorem does not prove the integrality of $\Gamma[u]$ in the general case.  

\par

However, besides the commutative case we can construct a simple example in which this integrality result holds. Let us define the three dimensional non commutative torus \cite{rieffel} as the algebra $\aa_{\theta}$ of power series of the form
\bbb
\mathop{\sum}\limits_{p_{1},p_{2},p_{3}\in\zzz} a_{p_{1},p_{2},p_{3}}
U_{1}^{p_{1}}U_{2}^{p_{2}}U_{3}^{p_{3}},
\eee
where $U_{1}$, $U_{2}$ and $U_{3}$ are unitary elements fulfilling the relations
\bbb
U_{i}U_{j}=e^{2i\pi\theta_{ij}}U_{j}U_{i},
\eee
with $\theta\in M_{3}(\rrr)$ an antisymmetric matrix. Moreover, we always assume that the sequence  $(a_{p_{1},p_{2},p_{3}})_{(p_{1},p_{2},p_{3})\in\zzz^{3}}$ decreases faster than any polynomial, which characterizes the analogue of "smooth functions" on the three dimensional noncommutative torus. 

\par

On $\aa_{\theta}$ we define a trace by
\bbb
\int\lp\mathop{\sum}\limits_{p_{1},p_{2},p_{3}\in\zzz} a_{p_{1},p_{2},p_{3}}
U_{1}^{p_{1}}U_{2}^{p_{2}}U_{3}^{p_{3}}\rp=a_{0,0,0},
\eee
which is completely similar to the usual integral since it singles out the constant mode in the Fourier expansion.

\par

Moreover, we define three derivations $\partial_{1}$, $\partial_{2}$ and $\partial_{3}$ by their actions on the generators
\bbb
\partial_{i}U_{j}=2i\pi\delta_{ij}U_{j},
\eee
where $\delta_{ij}$ equals 1 if $i=j$ and $0$ otherwise. These derivations are analogous to the derivations with respect to the standard coordinates on the usual torus.

\par

This construction yields a 3-dimensional spectral triple $(\aa,\hh,\dd)$, where $\aa$ is the algebra $\aa_{\theta}$ acting by multiplication on the Hilbert space $\ov{\aa_{\theta}}\ot\ccc^{2}$, where $\ov{\aa_{\theta}}$ is the completion of $\aa_{\theta}$ for the scalar product defined by the trace. The Dirac operator is $\dd=i\sigma_{\mu}\partial_{\mu}$, where $\sigma_{\mu}$, $\mu=1,2,3$ denote the Pauli matrices and, as usual, summation over repeated greek indices $\lambda,\mu,\nu,\dots$ ranging from 1 to 3 is self-understood.  
The choice of the Pauli matrices means that we take the analogue of the euclidean metric on the noncommutative torus but other {\it constant} metrics $g^{\mu\nu}$ may be taken. In this case, one shows that the Chern-Simons action is independent of $g^{\mu\nu}$.  

\par

For the sake of brevity we do not give here any detailed calculation and refer to \cite{the} for a more thorough account. Let us simply state that this sepctral triple fulfils the closedness condition and that the Chern-Simons action is
\bbb
S_{CS}[A_{\mu}]=\frac{k}{4\pi}\int \epsilon_{\lambda\mu\nu}\t\lp A_{\lambda}\partial_{\mu}A_{\nu}
+\frac{2}{3}A_{\lambda}A_{\mu}A_{\nu}\rp
\eee
where $A_{\mu}$ is a hermitian matrix with entries in $\aa_{\theta}$, $\epsilon_{\lambda\mu\nu}$ is the completely antisymmetric tensor with $\epsilon_{123}=1$ and $k\in\rrr$ is a coupling constant.

\par

Under the gauge transformation determined by the unitary $u$, we have $A_{\mu}\rightarrow uA_{\mu}u^{-1}+u\partial_{\mu}u^{-1}$, and the Chern-Simons action becomes
\bbb
S_{CS}[uA_{\mu}u^{-1}+u\partial_{\mu}u^{-1}]=S_{CS}[A_{\mu}]+\Gamma[u],
\eee  
with 
\bbb
\Gamma[u]=\frac{k}{12\pi}\int\epsilon_{\lambda\mu\nu}
\t\lp\partial_{\lambda}u\partial_{\mu}u^{-1}\partial_{\nu}u\rp.
\eee

\par

On the other hand, it is easily seen that the $(\aa,\hh,\dd)$ is a three dimensional spectral triple with simple dimension spectrum, so that one can apply the index theorem. For $\Re(z)>3/2$, the trace in the full Hilbert space $\hh\ot\ccc^{N}$ given by
\bbb
\t\lp udu^{-1} |\dd|^{2z}\rp=i\t\lp u\sigma_{\mu}\partial_{\mu}u^{-1} |\dd|^{2z}\rp
\eee
vanishes identically because it involves a trace on a single Pauli matrix. Accordingly, its residue is 0 and we have
\bbb
\tau_{0}\lp udu^{-1}|\dd|^{-1}\rp=0.
\eee
The same result holds for $\tau_{0}(u\nabla(du^{-1})|\dd|^{-1})$ and for $\tau_{0}(u\nabla^{2}(du^{-1})|\dd|^{-1})$, so that we have 
\bbb
\phi_{1}(u,u^{-1})=0.
\eee
Finally, let us compute 
\bbb
\phi_{3}(u,u^{-1},u,u^{-1})=\frac{1}{12}\mathop{\mathrm{res}}\limits_{z=0}
\t\lp u[\dd,u^{-1}][\dd,u][\dd,u^{-1}]|\dd|^{-3-2z}\rp.
\eee
Because of the relation $\sigma_{\lambda}\sigma_{\mu}\sigma_{\nu}=i\epsilon_{\lambda\mu\nu}$, the trace over Pauli matrices simply yields $2i\epsilon_{\lambda\mu\nu}$ and we get, bearing in mind that the scalar product on $\hh$ is given by $\int$,
\bbb
\t\lp udu^{-1}dudu^{-1}|\dd|^{-3-2z}\rp=
2\int\epsilon_{\lambda\mu\nu}
\t\lp u\partial_{\lambda}u^{-1}\partial_{\mu}u\partial_{\nu}u^{-1}\rp
\t\Delta^{-3/2-z},
\eee
where $\Delta$ denotes the standard 3-dimensional Laplacian on the commutative torus. It is important to notice that the LHS of the previous equation involves two different traces: the first trace denotes a trace on $M_{N}(\aa)$ whereas the second one is to be taken over all non zero modes of the Laplacian $\Delta$.  
\par

Using the relation 
\bbb
\mathop{\sum}\limits_{k\in\zzz} e^{-tk^{2}}
\mathop{\sim}\limits_{t\rightarrow 0}\sqrt{\frac{\pi}{t}},
\eee
we obtain (see \cite{gilkey} for a detailed account)
\bbb
\mathop{\mathrm{res}}\limits_{z=0}\t\lp\Delta^{-3/2-z}\rp=
\mathop{\mathrm{res}}\limits_{z=3/2}\t\lp\Delta^{-z}\rp=\frac{1}{4\pi^{2}}.
\eee
Gathering everything together, we get
\bbb
\phi_{3}(u,u^{-1},u,u^{-1})=
\frac{1}{24\pi^{2}}\int\epsilon_{\lambda\mu\nu}
\t\lp u\partial_{\lambda}u^{-1}\partial_{\mu}u\partial_{\nu}u^{-1}\rp.
\eee
Because $\phi_{1}$ vanishes identically, the index theorem shows that
\bbb
\frac{1}{24\pi^{2}}\int\epsilon_{\lambda\mu\nu}
\t\lp u\partial_{\lambda}u^{-1}\partial_{\mu}u\partial_{\nu}u^{-1}\rp\in\zzz,
\eee
which is completely analogous to the classical case, even for such a highly noncommutative "manifold". Accordingly, $\Gamma[u]$ belongs to $2i\pi\zzz$ as soon as $k\in\zzz$. This quantization of the coupling constant makes the partition function
\bbb
Z(\aa_{\theta})=\int[\dd A_{\mu}]\,e^{\frac{ik}{12\pi}S_{CS}[A_{\mu}]}
\eee
well defined after gauge fixing, where the measure $[\dd A_{\mu}]$ has to be understood as a product of all the one dimensional measures pertaining to the Fourier modes.

\par

To show that this index is actually non trivial, let us construct a simple example using the Power-Rieffel \cite{pacific}. Let $\theta$ be the deformation matrix given by
\bbb
\theta=\pp{0&\alpha&0\cr -\alpha&0&0\cr 0&0&0}\quad \alpha\in [0,1].
\eee  
In the algebra $\aa_{\theta}$, we can construct a hermitien projection $e$ by
\bbb
e=(U_{1}f(U_{2}))^{*}+g(U^{2})+U_{1}f(U_{2},
\eee
where $f_{1}$ and $f_{2}$ are suitable smooth functions on $S^{1}$, that we may always choose such that \cite{cras} $\int e=\theta$ and
\bbb
\frac{1}{2i\pi}\int e(\partial_{1}e\partial_{2}e-\partial_{2}e\partial_{1}e)=1.
\eee
{}From this projection, let us construct
\bbb
U=\frac{U_{3}+U_{3}^{*}}{2}+(2e-1)\frac{U_{3}-U_{3}^{*}}{2}.
\eee
Because $e$ is a hermitian projection, $U$ is unitary and we have
\bbbb
\partial_{1}U&=&2\partial_{1}e\frac{U_{3}-U_{3}^{*}}{2}\n\\
\partial_{2}U&=&2\partial_{2}e\frac{U_{3}-U_{3}^{*}}{2}\n\\
U\partial_{3}U^{-1}&=&2i\pi(2e-1).
\eeee
This yields, after a lengthy but straightforward computation,
\bbb
\int\epsilon_{\lambda\mu\nu}
\t\lp u\partial_{\lambda}u^{-1}\partial_{\mu}u\partial_{\nu}u^{-1}\rp=24\pi^{2}.
\eee
Furthermore, if we replace $U$ by $U^{n}$, we obtain an index equal to $n$.

\vskip 0.5truecm
\noindent

{\Large\bf Aknowledgements}\\

It is a pleasure to thank the C. Buzzanca and D. Kastler who have organized the ISI Guccia meeting "Quantum groups, noncommutative geometry and fundamental physical interactions" in Palermo in December 1998 and who gave me the opportunity to present this work. I am also indebted to  B. Iochum, T. Sch\"ucker, K. Valavane and R. Wulkenhaar for their unvaluable help.

\end{document}